\documentclass[proceedings, preprint]{rmaa}

\usepackage{paralist}

\usepackage{psfrag,color}

\SetYear{2011}
\SetConfTitle{Fourth Science Meeting with the GTC}

\title{The dependence of the Lyman$_{\alpha}$ Luminosity Function on redshift using SHARDS} 

\author{
  J.M. Rodr\'\i{}guez-Espinosa,\altaffilmark{1, 2} 
  O. G\'onzalez-Mart\'\i{}n,\altaffilmark{1, 2}
  J. A. L\'opez-Aguerri, \altaffilmark{1, 2}
  C. Mu\~noz-Tu\~n\'on, \altaffilmark{1, 2}
  P. G. P\'erez Gonz\'alez, \altaffilmark{3}
  and A. Cava, \altaffilmark{3}}

\altaffiltext{1}{Instituto de Astrof\'\i{}sica de Canarias,
  La Laguna, Tenerife, Spain (jmr.espinosa@iac.es)}

\altaffiltext{2}{Astrophysics Dept. U. of La Laguna, La Laguna, Tenerife, Spain}
\altaffiltext{3}{Astrophysics Dept., U. Complutense de Madrid, Madrid, Spain}

\shortauthor{Rodriguez-Espinosa \& et al.}
\shorttitle{RevMexAA(SC) Ly_{\alpha} LF vs. z}

\listofauthors{J.M. Rodr\'\i{}guez-Espinosa, O. G\'onzalez-Mart\'\i{}n,  A. L\'opez-Aguerri, C. Mu\~noz-Tu\~n\'on, P. G. P\'erez Gonz\'alez, \& A. Cava}
\indexauthor{Rodr\'iguez-Espinosa, J.M.}
\indexauthor{Gonz\'alez-Mart'i{}n, O.}
\indexauthor{L\'opez-Aguerri, A.}
\indexauthor{Mu\~noz-Tu\~n\'on, C.}
\indexauthor{P\'erez Gonz\'alez, P.}
\indexauthor{Cava, A.}

\abstract{We report in this work on a project aimed at determining Ly$_{\alpha}$ luminosity functions from z=3 to z=6. The project is based on the 
use of very deep photometry from the SHARDS Survey, in a set of 24 medium band filters in the GOODS-N field. We present here some preliminary work carried out with four test 
images in four consecutive bands. We use the narrow band selection technique for searching emission line candidates. Eleven candidates have been detected so far, many of which are strong Ly$_{\alpha}$ candidates. In particular,  we have seen a firm candidate to an interacting pair of Ly$_{\alpha}$ sources at z=5.4.}

\resumen{Presentamos resultados preliminares de un proyecto cuya meta es obtener funciones de luminosidad de los emisores Ly$_{\alpha}$ desde z = 3 hasta z = 6. Para ello
usaremos datos del proyecto SHARDS, que est\' a obteniendo fotometr\'ia profunda en 24 filtros de anchura media, en el campo de GOODS-N.  Mostramos aqu\'i  resultados
preliminares obtenidos con im\'agenes reducidas en cuatro de los filtros. Usamos la t\'ecnica de selecci\'on de fuentes con l\'ineas de emisi\'on mediante observaciones con filtros de banda estrecha. Se han detectado, hasta ahora, once candidatos en estos cuatro filtros, varios de los cuales son con seguridad fuentes Ly$_{\alpha}$. En particular, dos de los objetos son firmes candidatos a ser una pareja de emisores Ly$_{\alpha}$ a z = 5.4 en interacci\'on}

\addkeyword{Galaxies: high-redshift}
\addkeyword{Galaxies: luminosity function}

\begin{document}
\maketitle

\section{Introduction}
\label{sec:intro}
The Ly$_{\alpha}$ emission line and the rest-frame UV continuum are the best tracers of star formation within the optical-NIR window for galaxies up to the end of the
cosmic reionisation era at z$\sim 6$ (Fan et al 2006). Star forming galaxies at high-z have been detected either by narrow-band filter imaging, aimed at identifying strong 
Ly$_{\alpha}$ emitters (LAEs), or by rest-frame UV multi-band imaging, looking for the UV continuum break (LBGs) due to the Ly$_{\alpha}$ forest (e.g., Steidel et al. 2005; 
Gronwall et al. 2007). Since the  Ly$_{\alpha}$ escape fraction is so sensitive to the amount of neutral hydrogen present in the intergalactic medium, the relative abundance 
of LAEs and LBGs offer a powerful probe of both the cosmic SFR density and the fraction of neutral gas. However, most studies so far have focused on the most luminous LAEs, 
while there is mounting evidence that the faint end of the galaxy luminosity function steepens significantly for intrinsically lower luminosity objects at higher z, indicating 
that the bulk of the cosmic SFR density at z $\sim$ 6.5 is due to an abundant population of low luminosity star forming galaxies (Ouchi et al 2009; Oesch et al. 2009). Moreover,  
there is significant evidence showing that the number density of ionizing sources detected at z $\sim 6$ is not sufficient to account for the full ionization of the universe observed 
at z $\sim 6$. In fact recent observations show that very steep LFs are necessary to balance the re-ionization budget (Trenti et al. 2010). In this project we will try to address 
these questions.

\section{Methodology}
\label{sec:Method}
To answer the above questions we will use deep photometry in 24 medium band filters, kindly provided to us by the 
SHARDS collaboration, (PI Dr. P. G. P\'erez Gonz\'alez). Our aim is to detect  Ly$_{\alpha}$ emitting sources, both LAES and 
LBGs. For this detection we will employ the two techniques that have proven most successful in detecting high-z emission 
line sources, namely narrow band searches and continuum selection searches. The first technique relies on the comparison of
images in contiguous filters for objects visible only in one of the images. This method spots easily strong line emitters. 
The second technique consists in searching for objects that are seen in various consecutive red filters and eventually cease to 
be visible bluer than a given filter. These are  the so called drop out sources, and are based on the fact that  Ly$_{\alpha}$ continuum 
sources show a prominent rest frame UV continuum that is easily seen redder of the Ly$_{\alpha}$ line, but it is deeply absorbed 
bluer that this line. With the sources detected in each of the 24 filters we will determine the Ly$_{\alpha}$ volume density as a 
function of z. Furthermore, we intend to build up Ly$_{\alpha}$ luminosity functions at any z between z= 3 and z= 6.7. This 
will allow us to follow up the source number density from the epoch of reionisation till the epoch (z=3) in which the Ly$_{\alpha}$ source population was depressed with respect to the Lyman Break source population. 

\begin{figure}[!t]\centering
  \includegraphics[width=\columnwidth]{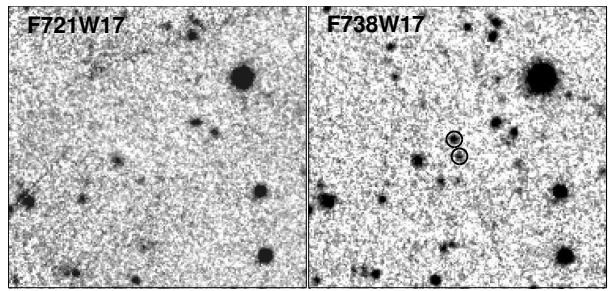}
  \caption{Composite of two images of the same part of the field observed in two adjacent SHARDS filters. The 
  black circles show two object tat are only seen in the F738 filter but not in the bluer filters. They are likely a pair of interacting galaxies at z$\sim 5$ }
  \label{fig:Fig1} 
\end{figure}

\section{Preliminary Results}
\label{Results}
So far we have been testing our method with SHARDS images  of the GOODS-N field in four different filters. The filters used are the F687W17, F704W17, F721W17 \& F738W17 respectively, where the names of the filters describe approximately the central wavelength and width (both in nm) for each filter. In this paper we show the results of the search carried out for objects appearing in filter F738W17 that are not seen in the bluer filters.  Once detected, the positions of the objects have been searched for in the other SHARDS filters, using a special flavor of the UCM Rainbow database Navigator (P\'erez-Gonz\'alez et al. 2008; Barro et al. 2011), as well as in Broad Band images of the same field. Figure 1 is a combination of a small section of this field as seen in two filters, namely the F721W17 and the F738W17. Black circles show the position of two sources seen in the F738W17 filter but not in any of the other 3 bluer filters. A quick examination of the sources detected, using the Rainbow data base , shows that at least 6 of these sources are {\it bona fide} Ly$_{\alpha}$ emitters. In particular, the two objects at z$\sim 5$, marked in Fig. 1, are within 10'' of each other. We need still to confirm that these two objects do have similar redshifts, in which case they would likely be a pair of interacting galaxies at z$\sim 5$. Figure 2 is a SED, built with data from the SHARDS filters, for one of these objects.  A conspicuous Ly$_{\alpha}$ line is readily seen. The photometric-z quoted is based on the wavelength of the filter in which the line shows prominently. A hint of other line is seen although its precise wavelength will have to wait for a more refined analysis.

\begin{figure}[!t]\centering
  \includegraphics[angle=-90, width=0.8\columnwidth]{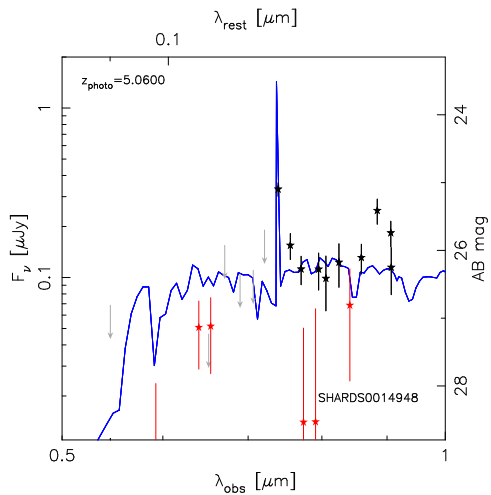}
  \caption{SED of object SHARDS0014948. The black points are SHARDS data, while the red points are broadband data. The gray arrows are upper limits. The blue line is a quick fit to a blue stellar population. Note that no Ly$_{\alpha}$ forest has been included in the population model. The gray upper limits already show that the continuum blue-wards of Ly$_{\alpha}$ is severely depressed, as expected for a LBG source.}
  \label{fig:Fig2}
\end{figure}

\begin{acknowledgements}
This work has made use of a private version of the Rainbow Cosmological Surveys Database, which is operated by the Universidad Complutense de Madrid (UCM). The authors are partially funded by the Consolider-Ingenio (CSD00070-2006) and Estallidos (AYA2010-21887-C04-04) grants from the Spanish MICINN.

\end{acknowledgements}

\end{document}